\documentclass[11pt]{article}

\usepackage[preprint]{acl}

\usepackage{times}
\usepackage{latexsym}
\usepackage{booktabs}
\usepackage[table]{xcolor}
\usepackage{hyperref}

\usepackage[T1]{fontenc}

\usepackage[utf8]{inputenc}

\usepackage{microtype}

\usepackage{inconsolata}

\usepackage{graphicx}
\usepackage{amsmath}
\usepackage{cleveref}
\usepackage{longtable}
\usepackage{multirow}

\usepackage{enumitem}

\title{Overreliance on AI in Information-seeking from Video Content}

\author{Anders Giovanni Møller \\
  IT University of Copenhagen \\
  \texttt{agmo@itu.dk} \\ \And
  Elisa Bassignana \\
  IT University of Copenhagen \\
  Bocconi University \\
  \texttt{elba@itu.dk} \\ \AND
  Francesco Pierri\textsuperscript{*} \\
  Politecnico di Milano \\
  \texttt{francesco.pierri@polimi.it} \\ \And
  Luca Maria Aiello\textsuperscript{*} \\
  IT University of Copenhagen \\
  Pioneer Centre for AI \\
  \texttt{luai@itu.dk} \\}

\begin{document}
\maketitle

\begin{abstract}

The ubiquity of multimedia content is reshaping online information spaces, particularly in social media environments. At the same time, search is being rapidly transformed by generative AI, with large language models (LLMs) routinely deployed as intermediaries between users and multimedia content to retrieve and summarize information.
Despite their growing influence, the impact of LLM inaccuracies and potential vulnerabilities on multimedia information-seeking tasks remains largely unexplored.
We investigate how generative AI affects accuracy, efficiency, and confidence in information retrieval from videos.
We conduct an experiment with around 900 participants on 8,000+ video-based information-seeking tasks, comparing behavior across three conditions: (1) access to videos only, (2) access to videos with LLM-based AI assistance, and (3) access to videos with a deceiving AI assistant designed to provide false answers. 
We find that AI assistance increases accuracy by 3--7\% when participants viewed the relevant video segment, and by 27--35\% when they did not. Efficiency increases by 10\% for short videos and 25\% for longer ones. However, participants tend to over-rely on AI outputs, resulting in accuracy drops of up to 32\% when interacting with the deceiving AI. Alarmingly, self-reported confidence in answers remains stable across all three conditions.
Our findings expose fundamental safety risks in AI-mediated video information retrieval.

\end{abstract}

\section{Introduction}

\renewcommand{\thefootnote}{\fnsymbol{footnote}}
\setcounter{footnote}{0}
\footnotetext{\textbf{\textsuperscript{*}}Francesco Pierri and Luca Maria Aiello equally supervised this work.}
\renewcommand{\thefootnote}{\arabic{footnote}}
The growing dominance of video-based multimedia content is reshaping online information spaces, as users increasingly turn to platforms like YouTube and TikTok for news, learning, and everyday information~\cite{pewresearch2025}.
Meanwhile, generative AI is changing how individuals search for and process information, with Large Language Models (LLMs) acting as intermediaries between users and the underlying sources~\cite{sadeddine2025large}. 
By mediating what information users encounter, how it is summarized, and which perspectives are emphasized or omitted, LLMs can actively shape users’ information environments~\cite{liu2023evaluating,miroyan2026search}.
Recent reports suggest that a large fraction of users' interactions with chatbots involve information-seeking queries~\cite{NBERw34255,bassignana-etal-2025-ai,savoldi2025generative}. Many of these systems are increasingly integrated directly into major online platforms—for example, AI assistants embedded in social media services such as Grok in X~\cite{renault2025grok}. Because these LLM-based systems are largely designed and deployed by a small number of private companies, the growing reliance on AI-mediated search may concentrate significant influence over how information is accessed and presented to users.
At the same time, AI-powered summarization and ``overview'' features are now embedded across major search engines, mediating an increasing share of online information access~\cite{li2025humantrustaisearch}. 
As a result, AI-mediated search is rapidly becoming a central interface for accessing online knowledge.

However, a fundamental challenge to this new paradigm is that LLMs can often provide incorrect outputs \cite{min-etal-2023-factscore,wang-etal-2024-factuality,augenstein2024factuality,hallucination} as well as amplify existing stereotypes and biases \cite{biasesllms,fang2024bias,gallegos-etal-2024-bias}. 
These limitations are particularly concerning in high-stakes settings where incorrect information may mislead and cause harm~\cite{bean2026reliability}, especially when users over-rely on AI tools~\cite{passi2025addressing}.

\begin{figure*}[!t]
    \centering
    \includegraphics[width=\textwidth]{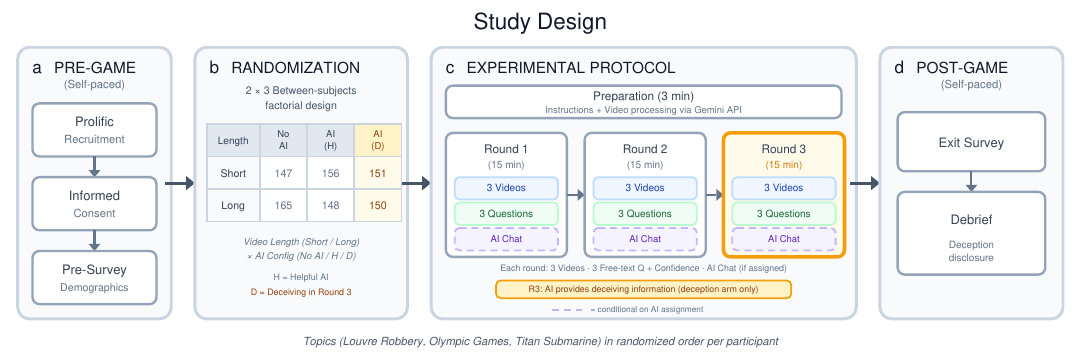}
    \caption{
        Overview of the study design. 
    }
    \label{fig:experiment}
\end{figure*}

\begin{figure}[ht]
    \centering
    \includegraphics[width=\columnwidth]{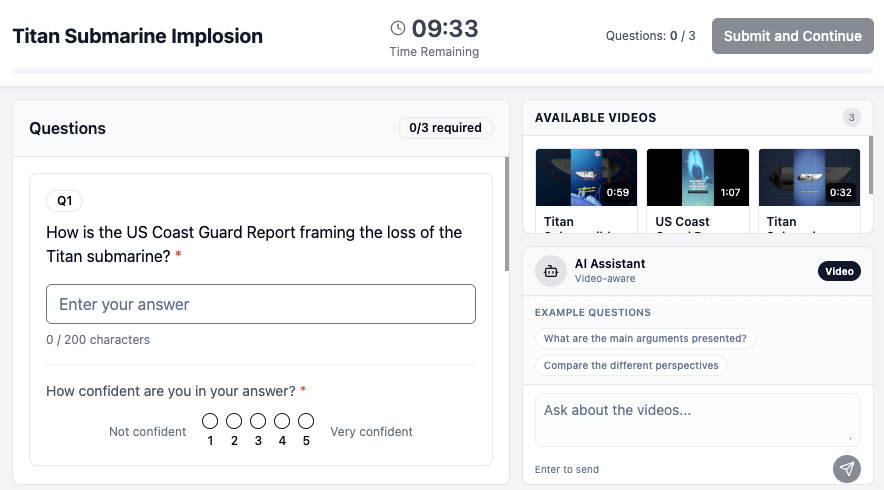}
    \caption{Platform overview for the short video condition with AI assistance. The left panel displays the questions, followed by an open-text field for answering and a Likert scale for expressing confidence in the answer. The top-right panel shows three video thumbnails, each of which opens a full-size video player when clicked. The bottom-right panel provides a chat interface where participants can ask questions to an AI assistant, which is clearly labeled as having access to the video content.}
    \label{fig:platform}
\end{figure}

As information seeking shifts toward online video and other multimedia~\cite{pewresearch2025}, we know little about how generative AI shapes information-seeking behavior when sources are multimodal.
LLM inaccuracies may be amplified when AI summarizes or interprets video, because videos are harder to verify than text, and users may be more inclined to skip verifying the source content. 
How users engage with AI-generated multimedia interpretations, and how this affects their interpretation of multimedia content, remain open questions.

Our research addresses this gap by investigating how the use of generative AI affects accuracy, efficiency, and confidence in video-based information-seeking tasks.
We conduct a large-scale user study involving $917$ participants completing $8,253$ information retrieval tasks over video content of varying lengths (short: \textless 1 minute, long: 1-5 minutes). 
We compare user behavior across three conditions: (1) access to videos only, (2) access to videos and a ``helpful'' LLM-based AI assistant, and (3) access to videos and a ``deceiving'' LLM-based AI assistant designed to provide incorrect answers in the final round of the task (see Figure~\ref{fig:experiment}). 
This setup allows us to quantify both the benefits of AI-assisted information-seeking and the risks associated with inaccurate or adversarial AI behavior. 
We aim to answer the following research questions: 

\noindent
\textbf{RQ1}: \textit{To what extent does AI assistance improve performance in information-seeking tasks involving video-based content?} 

\noindent
\textbf{RQ2}: \textit{To what extent does over-reliance on AI outputs affect performance when the system provides deceiving information?}

By analyzing accuracy, efficiency, and user confidence across our three conditions (no AI, helpful AI, and deceiving AI), our work provides the first systematic investigation of how generative AI shapes human information retrieval from video content. 
Our findings reveal that participants who could choose whether to watch the videos or use the AI improved answer accuracy with AI assistance by +3--7\% points when participants watched relevant video segments and by +27--35\% points when they did not. Efficiency, measured as task completion speed, increased by +10.5\% for short videos and +25\% for long ones. However, users tend to over-rely on AI outputs, leading to substantial accuracy drops when the system outputs false information, decreasing by -5--9\% points when participants watched the answer segment, and by -29--32\% points when they did not. 
Alarmingly, self-reported confidence in the given answers remains stable even when relying exclusively on AI to complete the information-seeking tasks (without having watched the video), exposing fundamental safety risks in AI-mediated video information retrieval.

Our contributions are:\footnote{We release our interface, data, and code: \href{https://github.com/AndersGiovanni/vllm-search/}{repository link}.}

\begin{itemize}[leftmargin=*]
    \itemsep0em
    \item We collect the first-of-its-kind dataset of how users process video content to complete information-seeking tasks, with and without AI assistance.
    \item Through a controlled experiment, we estimate the effect of having access to helpful vs deceiving AI assistance on accuracy, efficiency, and confidence in information retrieval from video content.
\end{itemize}

\section{Related Work}
Existing work on AI-assisted information-seeking has largely examined chatbots and conversational search systems that integrate retrieval. 

Much of this literature evaluates the quality, credibility, and diversity of sources surfaced or cited by LLMs.
For instance, \citet{dai2025media} show that citation bias in LLM-generated answers is driven more by which outlets models preferentially cite than by the ideological slant of the cited content itself.
\citet{park2025measuring} document systematic media-outlet-name bias in LLMs and propose measurement and mitigation methods that reduce this bias in downstream tasks.
\citet{minici2025auditing} audit LLM-mediated news exposure and find reduced outlet diversity and model-specific skews in the distribution of cited sources, consistent with distinct implicit ``editorial'' behaviors.

To support systematic comparisons, recent work introduced Search Arena~\cite{miroyan2026search}, a benchmark for LLM-based search that comprises over 24K paired multi-turn interactions with search-augmented LLMs (across diverse intents and languages), including full system traces and ~12K human preference votes.
An empirical analysis of these interactions reveals that user preferences are influenced by citation count and source type.

Researchers have also studied how LLM-powered search changes user behavior and downstream outcomes. 
For example, \citet{10.1145/3613904.3642459} conduct two experiments to test whether conversational search increases selective exposure relative to conventional search, and how opinion-biased LLM responses—either reinforcing or challenging a user’s views—moderate this effect. 
\citet{li2025humantrustaisearch} compare generative AI against traditional search on a representative sample of the US population. They find that while participants trust generative AI less than traditional search, reference links and citations significantly increase trust in AI, even when those links and citations are incorrect or hallucinated. Similarly, \citet{spatharioti2025} also examine how LLM-based search affects behavior compared to traditional search. Participants assigned to LLM-based search completed tasks more quickly, while maintaining comparable accuracy when the LLM was correct; however, users overrelied on incorrect information when the model erred. 
\citet{taylor2024ai} and \citet{luettgau2025conversationalaiincreasespolitical} experimentally compare LLM-based conversational search with traditional web search in political information-seeking. Across US and UK samples, they find that chatbots improve political knowledge relative to no assistance, but yield learning gains broadly comparable to Google Search, with some evidence that direct engagement with ChatGPT improves comprehension.

Despite these advances, to the best of our knowledge, prior work has not directly examined AI-assisted information seeking in \emph{multimedia} settings, e.g., when the underlying evidence is video on social-media platforms rather than primarily text or linked webpages. 
As a result, we still lack an understanding of how LLM-generated summaries or interpretations of video content shape users’ search strategies, verification practices, and confidence.

\section{Methods}

\subsection{Study Design}

Figure~\ref{fig:experiment} provides an overview of the study design. The study consists of three experimental rounds. In each round, participants are presented with three open-ended questions and given the opportunity to watch three YouTube videos containing the answers to those questions. Each question is designed to be answered by information provided in exactly one video. Questions for a given video are randomly drawn from a pool of three questions per video, while ensuring that the three questions seen by participants cover information from the beginning, middle, and end of their respective videos, for adequate temporal coverage. The videos cover three topics, shuffled randomly across rounds: (1) the 2025 Louvre robbery, (2) the history of the Olympic Games, and (3) the 2023 OceanGate Titan submarine disaster. After providing each answer in an open-text field, participants report their confidence on a 5-point Likert scale. For each topic, we select six videos, three short (\textless 1 minute) and three long (1--5 minutes). Table~\ref{tab:videos} in Appendix~\ref{app:videos} contains the full list of videos.

We employ a 2$\times$3 between-subjects factorial design crossing video length (short, long) with AI condition (no AI, helpful AI, deceiving AI), for a total of six experimental conditions participants are randomly assigned to. Depending on the condition, participants have access either to the videos only (control group; C), to the videos and a helpful AI assistant with access to the video content (treatment group 1; T1), or to the videos and an AI assistant that behaves helpfully in rounds 1 and 2 (as in T1) but provides plausible yet incorrect answers in round 3 (treatment group 2; T2). 
This design allows us to estimate both the benefits of AI assistance and the risks of over-reliance on AI.

The AI assistant is implemented using Google \texttt{gemini-3-flash} (accessed through its API), a multimodal model capable of processing the content and metadata of YouTube videos directly via URL in real time, giving it access to the same video content available to participants. Participants interact with the assistant through a standard chat interface and may ask any question they wish. In the helpful AI condition (T1), the assistant is designed to provide accurate answers throughout all three rounds. In the deceiving AI condition (T2), the assistant behaves identically to T1 in the first two rounds, but is prompted to provide plausible yet incorrect answers in the third round. The full list of prompts is reported in the Appendix.

We recruit a pool of 1,026 participants via Prolific, restricting participation to fluent English speakers from the US and UK. Figure~\ref{fig:platform} shows the custom-built interface that the participants see, implemented using the Empirica framework~\cite{Almaatouq_Becker_Houghton_Paton_Watts_Whiting_2021}.

\subsection{Experimental Workflow}

\noindent
\textbf{Pre-Game.} Participants first complete a self-paced pre-study where they provide informed consent and answer a short survey. 

\noindent
\textbf{Experimental rounds.} During a 3-minute preparation stage, participants are introduced to the task, platform, and available tools (see \Cref{app:instructions} for details). In the AI-assisted conditions, the videos for the upcoming topic are processed in the background via the Gemini API, enabling the chat interface to provide instant responses. In each round, participants must answer all questions and report their confidence in each answer before proceeding. Each round has a time cap of 15 minutes.

\noindent
\textbf{Post-Game.} After the three experimental rounds, participants complete a self-paced post-study phase consisting of an exit survey. This survey assesses social media usage, familiarity with and trust in AI systems, prior knowledge of the topic, and perceived content reliability. Lastly, participants are presented with a debriefing statement disclosing the possibility of deceiving AI-generated information. List of survey items is provided in \Cref{app:response-distribution}.

\subsection{Analysis}

\noindent
\textbf{Quality checks and filtering.} We apply multiple quality checks prior to analysis. First, we exclude partial submissions and retain only participants who completed the full study. Second, we exclude participants who failed the attention check in the post-study survey. After these two filters, the final sample consists of 917 participants, distributed across conditions (see Figure~\ref{fig:experiment}b for a breakdown). Third, we use experimental duration as an additional quality indicator and find that $96.6\%$ of participants spent at least an average of 5 minutes per question round. Finally, we manually reviewed a sample of 603 open-ended answers and found them all to be well-formed and meaningful, indicating that participants did not provide nonsensical responses to complete the task quickly.

\noindent
\textbf{Annotating answer correctness.} We evaluate the correctness of participants' open-ended answers using an LLM-as-a-judge pipeline~\cite{zheng2023judging}. The LLM is given the participant's answer together with the reference answer and returns a binary label indicating whether the response is correct. We use three frontier models, Claude Opus 4.5, Gemini 3 Pro, GPT-5.2 (all with default parameters accessed through their API), and aggregate their outputs using majority voting. We apply the same procedure, though with a different prompt, to verify that the deceiving assistant produces incorrect answers in round 3, accounting for over 90\% of responses. We report the AI assistant's accuracy per round in each experimental condition in Appendix \ref{app:aiaccuracy}. The list of prompts is in Appendix \ref{app:prompts}.

To validate the annotation pipeline, four authors independently annotated a sample of 603 participant answers. Each item is annotated by exactly two human annotators, with items distributed across annotator pairs. For this sample, we observe high human inter-annotator agreement (Fleiss' $\kappa = 0.866$). Similarly, the three LLM judges achieve high agreement (Fleiss' $\kappa = 0.891$). When comparing majority-vote human labels with majority-vote LLM labels, we reach near-perfect agreement (Cohen's $\kappa=0.910$).

\noindent
\textbf{Accuracy, confidence, and watched status.} Using the correctness annotations, we compute \emph{answer accuracy} as the proportion of correctly answered questions over the set of questions of interest, and \emph{answer confidence} as the mean self-reported confidence for those questions. We also determine whether participants viewed the video segment needed to answer each question before responding. We classify a segment as \emph{watched} when the participant’s viewing history overlaps with the annotated answer segment by more than 50\%, enabling a comparison of answer accuracy between \emph{watched} and \emph{not-watched} cases.

\noindent
\textbf{Regression analysis.} We estimate the effect of AI assistance on accuracy and confidence using ordinary least squares (OLS) regression. The unit of analysis is the participant-round observation, with each participant contributing three observations, one for each round, yielding a total of $N \times 3 = 2751$ samples. The deceiving AI condition in rounds 1 and 2 (where participants receive correct AI answers) is considered a helpful AI condition, with only round 3 observations used for deceiving AI coding. The regression model includes treatment conditions (Control as reference), video length (short as reference), and their interaction, reflecting the $3 \times 2$ factorial design. We also add fixed effects for round and topic and demographic controls for age group (18-29, 30-49, 50+), gender (woman vs. non-woman), education level (no bachelor's degree, bachelor's degree, graduate degree), and political affiliation (liberal, moderate, conservative, other). All categorical variables are dummy-coded, including reference values. The dependent variable (DV) denotes participant-round accuracy or confidence, fit as separate models.
$$
\begin{aligned}
\text{DV} \sim\; & \text{C(LLM config.)} \times \text{C(video length)} \\
                 & + \text{C(topic)}  + \text{C(age)} + \text{C(gender)} \\
                 & + \text{C(round)} + \text{C(education)} \\
                 & + \text{C(political affiliation)}
\end{aligned}
$$
\noindent
\textbf{Action sequences.} To characterize how participants approached each question, we analyze the action sequences prior to each answer. 
We define three basic event types: ``W'' watched the answer segment, ``V'' accessed the video but did not watch the answer segment, and ``AI'' used AI.
Based on the wall-clock timestamps, we delineate eight temporal sequence patterns capturing most observed behaviours: ``W'', ``V'', ``AI'', ``W $\rightarrow$ AI'', ``V $\rightarrow$ AI'', ``AI $\rightarrow$ W'', ``AI $\rightarrow$ V'', and ``Direct'' (i.e., directly answering the question without watching the video or using AI).

\section{Results}
\label{sec:results}

\subsection{Information-seeking metrics}

\begin{figure}[t!]
    \centering
    \includegraphics[width=\columnwidth]{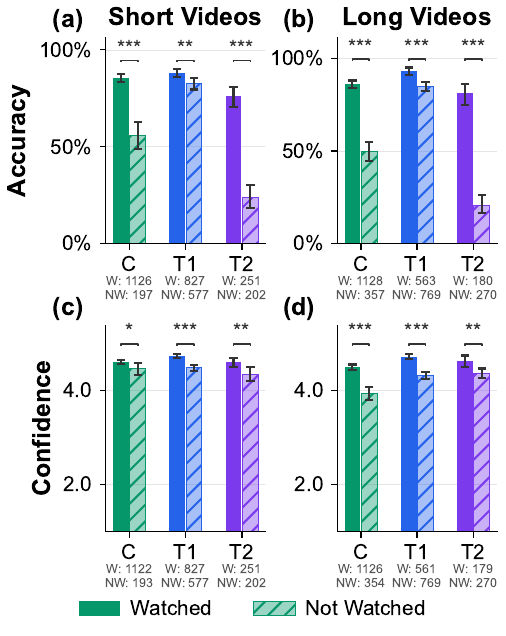}
    \caption{Accuracy and confidence of information-seeking tasks.
    The values for Watched (W) and Not Watched (NW) indicate whether the video segment containing the answer to the questions was watched or not. Within each condition, we compare accuracy using Fisher's exact test with Wilson score confidence intervals, and compare confidence using two-sided permutation tests ($10,000$ permutations) with bootstrap confidence intervals. Significance levels: *** $p<0.001$, ** $p<0.01$, * $p<0.05$}
    \label{fig:mainresults}
\end{figure}

\noindent \textbf{Accuracy and confidence.}
\Cref{fig:mainresults} reports the main results of our experiments in terms of accuracy and confidence across conditions. The analysis distinguishes between the helpful AI condition (T1) and the deceiving AI condition (T2) only in round 3, as this is the round in which the assistant provides wrong answers. Observations from rounds 1 and 2 in both AI-assisted conditions are represented in the T1 group. Results for each condition are further split according to whether answers were given after watching the video segments containing the information relevant to the question.

The results reveal three main findings. 
First, AI substantially improves answer accuracy when participants do not watch the video (\Cref{fig:mainresults}a,b). 
In the Control condition, accuracy drops considerably when participants answer without having watched the relevant video segment: an absolute decrease of -$29.7\%$ for short videos and -$36.6\%$ for long videos. When AI assistance is available (condition T1), this drop is reduced to only -$5.3\%$ and -$8.3\%$ for short and long videos, respectively. Given that approximately $30\%$ (short videos) to $40\%$ (long videos) of questions are answered without watching the relevant segment, AI can considerably improve the effectiveness of the information retrieval process. AI assistance also slightly improves accuracy for participants who did watch the video segments in the long video condition, with accuracy of $93.3\%$ in T1 vs. $86.2\%$ in Control.

\begin{figure*}[!t]
    \centering
    \includegraphics[width=\textwidth]{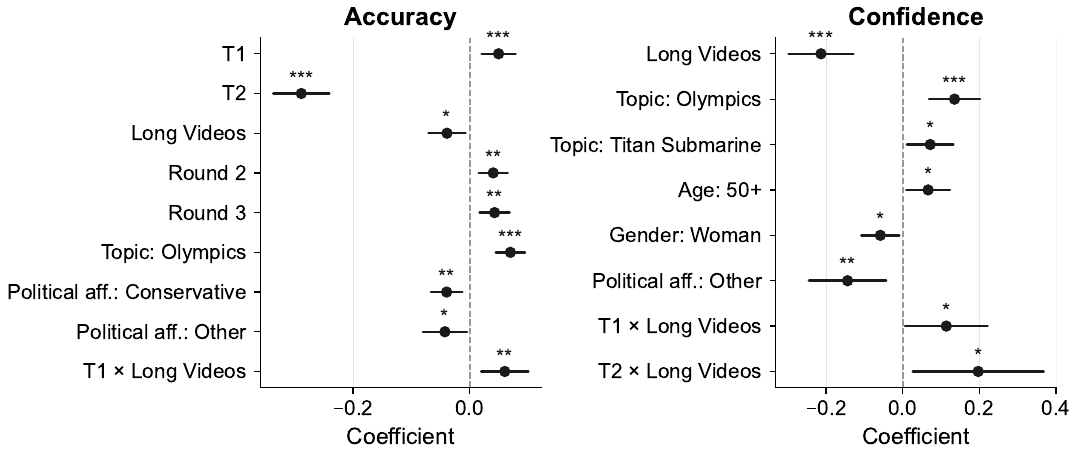}
    \caption{Significant coefficients of a regression model
    to predict answer accuracy (left) and answer confidence (right). Significance levels: *** $p<0.001$, ** $p<0.01$, * $p<0.05$}
    \label{fig:regression}
\end{figure*}

Second, participants appear to typically accept the answers provided by the deceiving AI, leading to a considerable decline in answer accuracy (\Cref{fig:mainresults}a,b). In the last round of condition T2, when the deceiving AI silently replaces the regular AI assistant, answer accuracy drops to around $20\%$ for questions answered by participants who did not watch the relevant video segments ($45\%$ of answers for short videos and $60\%$ for long videos). However, the accuracy drop compared to Control is much more contained when participants engage appropriately with the video content (-$9.4\%$ and -$5.1\%$ for short and long videos, respectively).

Third, and surprisingly, self-reported confidence in answers is relatively stable across all conditions (\Cref{fig:mainresults}c,d). We observe consistent yet modest confidence drops when participants did not watch the relevant video segments, but virtually no difference in confidence depending on whether AI assistance was available or not. Overall, average confidence remains high, with average values around 4.5 out of 5 and stable across rounds within each condition (\Cref{fig:round-confidence} in Appendix).

Taken together, these results reveal an alarming over-reliance on AI outputs for extracting information from video content. This is particularly evident in the T2 condition for the Not Watched group, where low answer accuracy stands in stark contrast with undiminished response confidence.

\noindent
\textbf{Regression results.}
\Cref{fig:regression}a shows the significant coefficients from the regression model to predict answer accuracy (the complete table with all coefficients is provided in the Appendix). The effect of AI on accuracy holds after controlling for all available covariates. The deceiving AI has the strongest negative coefficient of all predictors. The helpful AI has a positive effect on accuracy, but primarily because it is effective when used to gather information about long videos. This is consistent with the finding that the long-video condition alone negatively affects accuracy, albeit to a small extent. The wider the information space, the greater the challenge of identifying specific details~\cite{pirolli1999information}, and this challenge is mitigated by helpful AI assistance.

Beyond video length, accuracy is affected by two other factors related to the experimental setup. First, accuracy increases in rounds 2 and 3, likely as participants become more familiar with the interface and the task. Second, participants answer questions about the Olympics topic more accurately, suggesting that the corresponding questions may be slightly easier. Since participants encounter all topics in random order across the three rounds, differences in topic difficulty are balanced by the within-subject design and do not affect the main conclusions regarding AI's impact on the task.

Among the socio-demographic variables, those identifying as ``Conservative'' or ``Other'' (i.e., not ``Liberal'' nor ``Moderate'') show a negative effect on accuracy, suggesting slightly lower task performance. This pattern may be in line with prior studies indicating that conservatives are less accurate at distinguishing true from false statements in headlines and political claims~\cite{garrett2021conservatives,rathje2023accuracy}.

\Cref{fig:regression}b shows the regression results with confidence as the outcome variable. Consistent with the accuracy results, longer videos increase task difficulty and reduce participants' confidence in their answers. As before, the challenge posed by long videos is mitigated by access to AI assistance (T1-long and T2-long conditions). The Olympics topic also shows a positive effect on confidence, as does the 2023 OceanGate Titan disaster. In the US and the UK, where our participants are based, these two topics have likely received more media coverage than the 2025 Louvre robbery, potentially increasing participants' familiarity and, consequently, their confidence. Finally, consistent with prior work, we observe higher confidence levels among individuals aged 50+~\cite{orth2018development} and lower confidence levels among women~\cite{mccarty1986effects,10.1145/3510460}.

\subsection{Time efficiency}

\begin{figure}[t!]
    \centering
    \includegraphics[width=\columnwidth]{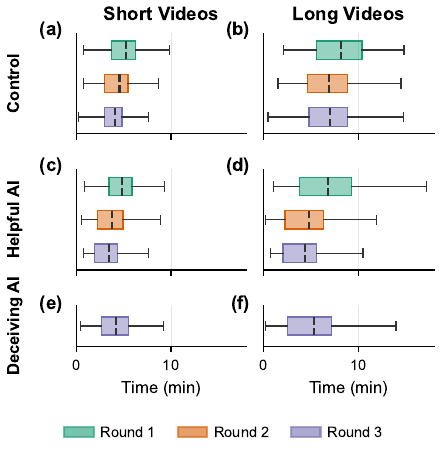}
    \caption{Distribution of time spent by participants across conditions and rounds. The black vertical lines in the boxes represent the average values.}
    \label{fig:time-spent-density}
\end{figure}

\Cref{fig:time-spent-density} reports the distribution of time spent by participants across the three rounds of information-seeking. As expected, participants assigned to the short video condition spent less time on the task than those assigned to the long video condition.
Additionally, as participants become more familiar with the task, their response times decrease in rounds 2 and 3 relative to round 1. 
To measure the efficiency gains from using AI assistance to extract information from video content, we compare the control condition (access to video only) against the experimental conditions with additional access to AI assistance (helpful or deceiving), across both video length conditions.
We observe average efficiency gains of +7.8\%, +16.8\%, +15.1\% in round 1, 2, 3, respectively, for the short video setup and of +16.5\%, +30.2\%, +37.1\% in the long video setup. Consistent with prior work on human-AI collaboration, access to AI assistance improves task efficiency by reducing task completion time~\cite{10.1093/qje/qjae044}. The efficiency gains are larger for tasks involving longer videos. 
Instead, we observe a -1.5\% slowdown relative to the control group in the short-video setup and a smaller efficiency gain of +24.1\% in the long-video setup when participants interact with the deceiving AI (round 3). This suggests that some participants may suspect the AI outputs and therefore spend additional time verifying them, reducing the overall efficiency gains.

\subsection{Behavioral patterns}

\begin{figure}[t!]
    \centering
    \includegraphics[width=\columnwidth]{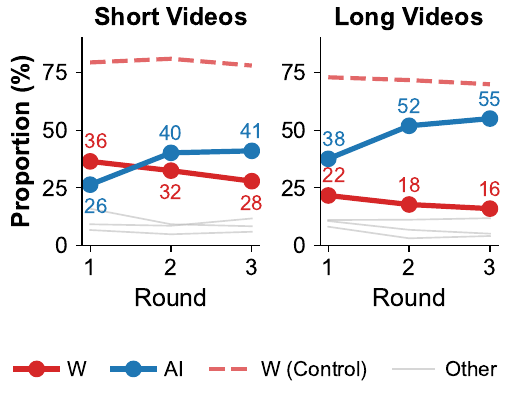}
    \caption{Behavioral patterns over the 3 rounds. 
    W: watched the video segment with the answer to the information-seeking question, AI: directly used AI to answer the question.
    Solid lines represent the treatment groups (T1 and T2), whereas the dashed line represents the control group. 
    Grey lines in the background represent other behavioral patterns that we omit for clarity, as they are much less prevalent.}
    \label{fig:sequence}
\end{figure}

\Cref{fig:sequence} reports the proportions of different action sequences performed by participants during the task. The figure groups all AI-assisted observations and focuses on the two most frequent contrasting approaches: ``W'': watching the video segment containing the answer before responding (without querying the AI), and ``AI'': directly querying the AI assistant without watching the relevant video segment. We observe that as rounds progress, participants tend to watch less video content: respectively $10\%$ and $6\%$ fewer participants do so from round 1 to round 3 for short and long videos. Conversely, reliance on AI increases: $15\%$ and $17\%$ more participants query the AI for short and long videos, respectively. The gap between the two behavioral patterns ``W'' and ``AI'' is larger for long videos, where participants are more likely to seek AI support to process the greater volume of information.

\begin{table}[t]
\centering
\scriptsize
\setlength{\tabcolsep}{4pt}
\begin{tabular}{lcccccccc}
\toprule
Group & W & V & \shortstack{W$\to$\\AI} & \shortstack{V$\to$\\AI} & \shortstack{AI$\to$\\W} & \shortstack{AI$\to$\\V} & AI & Direct \\
\midrule
\multirow{2}{*}{AI}
& 25.5 & 1.8 & 9.4 & 3.2 & 10.6 & 2.1 & 41.9 & 5.5 \\
& (1388) & (99) & (511) & (175) & (577) & (112) & (2283) & (300) \\
\addlinespace[2pt]
\multirow{2}{*}{Control}
& 75.2 & 7.7 & 0.0 & 0.0 & 0.0 & 0.0 & 0.0 & 17.1 \\
& (2113) & (215) & (0) & (0) & (0) & (0) & (0) & (480) \\
\bottomrule
\end{tabular}
\caption{Participant action sequences in the AI-assisted and control conditions.
Values are percentages, with counts in parentheses.}
\label{tab:sequence_patterns_main}
\end{table}

All other action sequences (grey lines in \Cref{fig:sequence}) are rare, with frequencies below $18\%$ (see Table~\ref{tab:sequence_patterns_main} and Table~\ref{tab:sequence_patterns_appendix} in the Appendix). Notably, the action sequence of querying the AI and subsequently fact-checking against the video content increases from $7.9\%$ in T1 (helpful AI) to $15.7\%$ in T2 (deceiving AI) for short videos and from $9.9\%$ in T1 to $13.8\%$ in T2 for long videos, indicating that a minority of participants critically evaluate AI outputs and question their accuracy when an answer seems implausible. Nevertheless, the overall trend points to a growing risk of over-reliance on AI over time.

\section{Discussion and Conclusion}

With recent progress in generative AI, the search paradigm has increasingly shifted from traditional search engines toward LLM-mediated information access~\cite{aral2026riseaisearchimplications}. However, LLMs have well-documented limitations, including hallucinations and biased or stereotyped outputs~\cite{li2025humantrustaisearch}. While these systems offer substantial benefits and efficiency gains across a variety of tasks~\cite{Noy2023ExperimentalEOA}, their current limitations raise concerns about their reliability as tools for information retrieval.

The highest risks arise when users rely entirely on LLMs to process information. In such settings, hallucinated, biased, or deliberately manipulated outputs can mislead users' beliefs or actions. This phenomenon is not marginal: an OpenAI report by~\citet{NBERw34255} finds that, across 1.1 million sampled conversations, $24.4\%$ of user interactions with ChatGPT are information-seeking queries, making this the second most common use case after ``practical guidance'' (i.e., how-to advice, at $28.8\%$). As AI assistants increasingly mediate information access, understanding how users interact with these systems and the potential risks that arise from such interactions becomes critical.

To advance our understanding of these new information retrieval practices, we conducted a controlled experiment investigating the effect of generative AI assistance on video-based information-seeking. We answer two research questions.

\noindent
\textbf{RQ1}: \textit{To what extent does AI assistance improve performance in information-seeking tasks involving video-based content?}
We find that helpful AI assistance improves participants' performance in both accuracy (up to $+35\%$ points) and time efficiency (up to $+25\%$  time saved). These gains are most pronounced as the information space grows larger, since participants are less inclined to browse long videos to locate target information and instead tend to rely directly on AI support.

\noindent
\textbf{RQ2}: \textit{To what extent does over-reliance on AI outputs affect performance when the system provides deceiving information?}
When the AI returns false yet plausible answers to information-seeking queries, participants typically accept the AI output without verifying it against the original video content (up to $51.3\%$ of participants in the long video condition). This behavior leads to substantial accuracy drops in the deceiving AI condition (-$29.1\%$ points and -$32.5\%$ points for short and long videos, respectively), driven largely by participants skipping the video and relying entirely on the AI response.
Only a minority of participants ($7.9$--$15.8\%$ across treatments) engaged in a more critical strategy by querying the AI and then fact-checking its response against the video.
Notably, participants' confidence remains largely unchanged across conditions (averaging 4.5 out of 5), suggesting that users struggle to detect when AI-generated outputs are incorrect.

Together, these findings highlight a fundamental tension in AI-mediated information access. While generative AI can substantially improve efficiency and performance, it can also foster over-reliance on system outputs, leaving users vulnerable to errors or manipulation~\cite{passi2025addressing}. We emphasize that our findings reflect the interaction between trust formed through prior helpful interactions and later-stage unreliability, rather than a general effect of exposure to a deceptive assistant. As AI assistants become increasingly integrated into search and information retrieval workflows, designing systems that encourage appropriate levels of trust, critical verification, and transparency will be essential to mitigate these risks.

\section*{Limitations}

Our study is limited to U.S. and U.K.-based crowdworkers on the Prolific platform and may not be representative of the broader population~\cite{pauketatprolific}. 
Given crowdworkers' familiarity with technology, they may be more likely to adopt and rely on AI assistance in order to complete tasks than the general population. 
Crowdworkers may use external LLMs to complete our study, despite platform policies against the use of LLMs. 

From an experimental setup perspective, our study is limited to a single frontier proprietary LLM. 
However, we observe that the LLM consistently outputs correct/deceiving information in our helpful AI and deceiving AI setups, respectively, which is the main purpose of our study.
In addition, we manually selected three topics, picked six videos for each topic (three short and three long), and designed three questions for each video.
While this design allows for controlled comparisons across conditions, it may limit the diversity of information-seeking scenarios and the generalizability of our findings to other topics, question types, or video domains.

\section*{Ethical Considerations}

The study was approved by the ethics board of the IT University of Copenhagen (internal approval case no. 2023-1767-18205003).
Crowdworkers were compensated for their time in accordance with the platform's recommendation of USD 16.28/hour. They may withdraw from the study by contacting the researchers. Subjects were fully anonymized in compliance with the GDPR and could opt out of providing sensitive sociodemographic information.
At the end of the study, subjects were informed about the presence of a deceiving AI assistant that may have provided incorrect information.
We used AI-based assistants to improve the clarity and language of the manuscript and to assist with aspects of the experimental setup, including code development and visualizations. All research ideas, methodological decisions, analyses, and interpretations, however, are solely those of the authors.

\section*{Acknowledgements}
Elisa Bassignana is supported by a research grant (VIL59826) from VILLUM FONDEN. Luca Maria Aiello acknowledges the support from the Carlsberg Foundation through the COCOONS project (CF21-0432). Francesco Pierri is partially supported by PNRR-PE-AI FAIR project funded by the NextGeneration EU program. 

\bibliography{custom, anthology-1, anthology-2}

\appendix

\section{Instructions to participants}
\label{app:instructions}

\begin{center}
\textbf{\large Study Instructions}\\[4pt]
\textit{Please read the following instructions carefully}
\end{center}

\subsubsection*{1\quad What You Will Do}

Your main task in this study is to \textbf{answer questions about three
different news events}. You will go through each event one at a time. You are
free to answer these questions however you prefer.

To help you find the information you need, you will have access to:
\begin{itemize}
    \item \textbf{Videos:} A collection of news clips about the event. You can
    watch as many or as few as you like, in any order.
    \item \textbf{AI Assistant:}\footnotemark{} An AI tool that has analyzed
    the videos. You can ask it questions to help you find information quickly.
\end{itemize}

\footnotetext{This item is only shown to participants in the AI treatment
conditions. Control-condition participants see only the Videos item.}

You may use these tools as much or as little as you want to complete the task.
Once you have answered the questions, you can continue.

\subsubsection*{2\quad Estimated Time}

This study will take approximately \textbf{20 minutes} to complete.

\noindent
\textit{Button:} \,
\fbox{\strut Start Study}

\section{Video Details}
\label{app:videos}

All videos were publicly available on YouTube at the time of data collection and are accessible at \url{https://youtu.be/<ID>}.                

  \begin{table}[ht]
  \centering
  \scriptsize
  \setlength{\tabcolsep}{3pt}
  \begin{tabular}{@{}clrc@{}}
  \toprule
  \# & Title & Dur. & YouTube ID \\
  \midrule
  \multicolumn{4}{@{}l}{\textit{Louvre Robbery (2025)}} \\
  \rowcolor[gray]{0.93}
  1 & Louvre heist masterminds' escape on camera & 0:58 & \texttt{xHC1hq7XDjY} \\
  2 & What crime scene footage reveals about heist & 0:54 & \texttt{4UFMKaLg-Fg} \\
  \rowcolor[gray]{0.93}
  3 & What to know about the suspects & 0:45 & \texttt{IO8JgFMz6e0} \\
  4 & How the Louvre Jewelry Heist Unfolded & 2:05 & \texttt{xLf1EneAyLQ} \\
  \rowcolor[gray]{0.93}
  5 & Paris Louvre heist: Jewel thieves escaping & 2:31 & \texttt{h4Adz7ydeno} \\
  6 & How Thieves Stole `Priceless' Jewels (WSJ) & 2:02 & \texttt{SASDkIQjouI} \\
  \addlinespace[3pt]
  \multicolumn{4}{@{}l}{\textit{Olympics}} \\
  \rowcolor[gray]{0.93}
  7 & Olympic Rings and Paris 2024 & 0:59 & \texttt{sUgqH1IXQQw} \\
  8 & History of the Olympics & 0:54 & \texttt{TTLCPcZilfM} \\
  \rowcolor[gray]{0.93}
  9 & Olympic Firsts and Facts & 0:59 & \texttt{sy7lGjteyRc} \\
  10 & Ancient vs Modern Olympics & 3:29 & \texttt{VdHHus8IgYA} \\
  \rowcolor[gray]{0.93}
  11 & Olympic Medals and Broadcasts & 3:10 & \texttt{fOyO6l75GU8} \\
  12 & Winter Olympics and Records & 1:28 & \texttt{zZkunt4BLEk} \\
  \addlinespace[3pt]
  \multicolumn{4}{@{}l}{\textit{Titan Submarine Implosion}} \\
  \rowcolor[gray]{0.93}
  13 & Titan Submarine: Timeline of the Implosion & 0:32 & \texttt{6DEqpCajJR8} \\
  14 & Titan Submersible: Key Facts and Figures & 0:59 & \texttt{AtQSd2NyZk0} \\
  \rowcolor[gray]{0.93}
  15 & US Coast Guard Report on Titan Submarine & 1:07 & \texttt{IeDqtdhX5IM} \\
  16 & The Titan Submersible: What Went Wrong? & 3:16 & \texttt{w9Q5WrRkefg} \\
  \rowcolor[gray]{0.93}
  17 & James Cameron on Titan Sub.\ Disaster & 3:37 & \texttt{LEBCc-Qpilw} \\
  18 & Inside the Titan Submersible & 2:49 & \texttt{pIeaCQPn53Y} \\
  \bottomrule
  \end{tabular}
  \vspace{0.3em}
  \caption{Videos}
  \label{tab:videos}
  \end{table}

  \section{Videos Questions}

Below we outline the 54 comprehension questions organized by topic and video. Each video has three questions, one for each of the temporal segments (beginning (B), middle (M), and end (E)). Expected open-text answers are listed in \textit{italic}. 

    \medskip
\noindent\textbf{Louvre Robbery (2025)}

\smallskip\noindent\textit{V1: Louvre heist masterminds' escape on camera}
\begin{itemize}[nosep,leftmargin=1em]
  \item[\textsc{b}] In which gallery did the thieves force their way in? --- \textit{Apollo gallery}.
  \item[\textsc{m}] On which floor did the robbery happen? --- \textit{First floor}.
  \item[\textsc{e}] What did the thieves drop? --- \textit{Empress Eug\'{e}nie's diadem}.
\end{itemize}

\smallskip\noindent\textit{V2: What crime scene footage reveals about heist}
\begin{itemize}[nosep,leftmargin=1em]
  \item[\textsc{b}] Which side of the Louvre did they enter from? --- \textit{Seine-facing side}.
  \item[\textsc{m}] What may be used together with an angle grinder? --- \textit{A blowtorch}.
  \item[\textsc{e}] What is the open investigation about? --- \textit{Aggravated theft and criminal conspiracy}.
\end{itemize}

\smallskip\noindent\textit{V3: What to know about the suspects}
\begin{itemize}[nosep,leftmargin=1em]
  \item[\textsc{b}] How many suspects have been handed preliminary charges? --- \textit{Four}.
  \item[\textsc{m}] How long after the heist was the first suspect arrested? --- \textit{Six days}.
  \item[\textsc{e}] How old is the man who arrived at the Louvre with a lift truck? --- \textit{37}.
\end{itemize}

\smallskip\noindent\textit{V4: How the Louvre Jewelry Heist Unfolded}
\begin{itemize}[nosep,leftmargin=1em]
  \item[\textsc{b}] How long after the museum opened did the heist happen? --- \textit{30 minutes}.
  \item[\textsc{m}] What percentage of the rooms have security cameras? --- \textit{75\%}.
  \item[\textsc{e}] What was the Louvre before being a museum? --- \textit{A palace}.
\end{itemize}

\smallskip\noindent\textit{V5: Paris Louvre heist: Jewel thieves escaping}
\begin{itemize}[nosep,leftmargin=1em]
  \item[\textsc{b}] How did the two thieves make their way down? --- \textit{With a boom lift}.
  \item[\textsc{m}] How long did it take the security guards to realize it was a robbery? --- \textit{Immediately}.
  \item[\textsc{e}] Why did the two security guards back away? --- \textit{Afraid the thieves might be armed}.
\end{itemize}

\smallskip\noindent\textit{V6: How Thieves Stole `Priceless' Jewels (WSJ)}
\begin{itemize}[nosep,leftmargin=1em]
  \item[\textsc{b}] In which gallery did the robbery take place? --- \textit{Apollo Gallery}.
  \item[\textsc{m}] When was the Mona Lisa stolen? --- \textit{1911}.
  \item[\textsc{e}] What could happen to the stolen pieces since they'd be hard to sell? --- \textit{Could be dismantled}.
\end{itemize}

\medskip
\noindent\textbf{Olympics}

\smallskip\noindent\textit{V7: Olympic Rings and Paris 2024}
\begin{itemize}[nosep,leftmargin=1em]
  \item[\textsc{b}] What does the interlinking of the Olympic rings represent? --- \textit{International cooperation}.
  \item[\textsc{m}] On which famous mountain has the 2024 Olympic torch traveled? --- \textit{Mount Everest}.
  \item[\textsc{e}] How many times did Paris host the Olympics? --- \textit{3}.
\end{itemize}

\smallskip\noindent\textit{V8: History of the Olympics}
\begin{itemize}[nosep,leftmargin=1em]
  \item[\textsc{b}] Who were the ancient Olympic Games dedicated to? --- \textit{God Zeus}.
  \item[\textsc{m}] Who founded the modern Olympic Games? --- \textit{Baron Pierre de Coubertin}.
  \item[\textsc{e}] How many times have the Olympic Games been canceled? --- \textit{3}.
\end{itemize}

\smallskip\noindent\textit{V9: Olympic Firsts and Facts}
\begin{itemize}[nosep,leftmargin=1em]
  \item[\textsc{b}] Who was allowed to participate in the first recorded Olympic Games? --- \textit{Freeborn Greek men}.
  \item[\textsc{m}] When was the Olympic flame relay introduced? --- \textit{1936 Berlin Games}.
  \item[\textsc{e}] How many runners finished the 1904 Olympic marathon? --- \textit{14}.
\end{itemize}

\smallskip\noindent\textit{V10: Ancient vs Modern Olympics}
\begin{itemize}[nosep,leftmargin=1em]
  \item[\textsc{b}] What does ``olympiads'' indicate in ancient Greek? --- \textit{4-year increment}.
  \item[\textsc{m}] What is the name of the 776 BC champion? --- \textit{Coroebus}.
  \item[\textsc{e}] Where did the 1896 Olympics take place? --- \textit{Athens}.
\end{itemize}

\smallskip\noindent\textit{V11: Olympic Medals and Broadcasts}
\begin{itemize}[nosep,leftmargin=1em]
  \item[\textsc{b}] When was the last solid gold medal handed out? --- \textit{1912}.
  \item[\textsc{m}] When were the Olympics broadcast on television for the first time? --- \textit{1936}.
  \item[\textsc{e}] How many Olympics were cancelled because of World War 2? --- \textit{2}.
\end{itemize}

\smallskip\noindent\textit{V12: Winter Olympics and Records}
\begin{itemize}[nosep,leftmargin=1em]
  \item[\textsc{b}] How many countries won medals in the first Winter Games (1924)? --- \textit{11}.
  \item[\textsc{m}] What happened to the torch for the first time in Sochi 2014? --- \textit{It went into space}.
  \item[\textsc{e}] Who was the first person to win 10 Olympic medals? --- \textit{Raisa Smetanina}.
\end{itemize}

\medskip
\noindent\textbf{Titan Submarine Implosion}

\smallskip\noindent\textit{V13: Titan Submarine: Timeline of the Implosion}
\begin{itemize}[nosep,leftmargin=1em]
  \item[\textsc{b}] How many passengers were in the Titan submarine? --- \textit{5}.
  \item[\textsc{m}] How many seconds does it take our brain to process pain? --- \textit{100 milliseconds}.
  \item[\textsc{e}] How many milliseconds did it take for Titan to implode? --- \textit{1 millisecond}.
\end{itemize}

\smallskip\noindent\textit{V14: Titan Submersible: Key Facts and Figures}
\begin{itemize}[nosep,leftmargin=1em]
  \item[\textsc{b}] After how much time did Titan lose connection with the mother ship? --- \textit{1 hr 45 minutes}.
  \item[\textsc{m}] How big was the water pressure at 3,500\,m depth? --- \textit{351 kg per square centimeter}.
  \item[\textsc{e}] When was debris from the wreckage collected? --- \textit{June 28}.
\end{itemize}

\smallskip\noindent\textit{V15: US Coast Guard Report on Titan Submarine}
\begin{itemize}[nosep,leftmargin=1em]
  \item[\textsc{b}] How is the report framing the loss of the Titan submarine? --- \textit{A preventable disaster}.
  \item[\textsc{m}] What were the primary contributing factors to the disaster? --- \textit{Inadequate design, certification, maintenance, and inspection}.
  \item[\textsc{e}] What did OceanGate do following the 2022 safety issues? --- \textit{Nothing}.
\end{itemize}

\smallskip\noindent\textit{V16: The Titan Submersible: What Went Wrong?}
\begin{itemize}[nosep,leftmargin=1em]
  \item[\textsc{b}] What was the Titan made of? --- \textit{Carbon fiber}.
  \item[\textsc{m}] Who was terrified to get into the submarine? --- \textit{Suleiman Dawood}.
  \item[\textsc{e}] What lawsuits are likely to be successful? --- \textit{Wrongful death and negligence}.
\end{itemize}

\smallskip\noindent\textit{V17: James Cameron on Titan Sub. Disaster}
\begin{itemize}[nosep,leftmargin=1em]
  \item[\textsc{b}] Who asked Cameron to go diving? --- \textit{Stockton Rush}.
  \item[\textsc{m}] How much time did it take to get confirmation about the loud bang? --- \textit{One hour}.
  \item[\textsc{e}] What is progressive failure with microscopic water ingress called? --- \textit{Cycling fatigue}.
\end{itemize}

\smallskip\noindent\textit{V18: Inside the Titan Submersible}
\begin{itemize}[nosep,leftmargin=1em]
  \item[\textsc{b}] What is Brian's job? --- \textit{Director of photography}.
  \item[\textsc{m}] What was the only way to get in or out of the Titan? --- \textit{Through the front}.
  \item[\textsc{e}] How many bolts go around the door? --- \textit{Four}.
\end{itemize}

\section{Surveys}

\subsection*{Pre-study Survey}

\begin{enumerate}
  \item \textbf{Age}: ``What is your age?'' \\
    \textit{Number input}.

  \item \textbf{Gender}: ``What is your gender?'' \\
    \textit{Options:} Man, Woman, Non-binary, Other, Prefer not to say

  \item \textbf{Education}: ``What is your highest level of education completed?'' \\
    \textit{Options:} Less than high school, High school diploma or equivalent, Some college, no degree, Associate's degree, Bachelor's degree, Master's degree, Doctorate or professional degree

  \item \textbf{Political Affiliation}: ``What is your political affiliation?'' \\
    \textit{Options:} Very Liberal, Liberal, Moderate, Conservative, Very Conservative, Other, Prefer not to say

  \item \textbf{Video Content Engagement}: ``To what degree do you prefer video-based content as opposed to image or text-based content?'' \\
    \textit{5-point Likert:} 1 = ``Mostly images/text'' to 5 = ``Mostly videos''
\end{enumerate}

\subsection*{Post-study Survey}

\begin{enumerate}
  \item \textbf{Social Media Daily Usage}: ``On average, how much time do you spend on social media per day?'' \\
    \textit{Options:} I don't use social media, Less than 1 hour, 1--2 hours, 2--4 hours, 4--6 hours, More than 6 hours

  \item \textbf{Social Media Platforms}: ``Which social media platforms do you use on a regular basis? (Select all that apply)'' \\
    \textit{Options:} TikTok, Instagram, YouTube, Facebook, Twitter/X, Reddit, Snapchat, LinkedIn, Bluesky, Other, None

  \item \textbf{AI Usage Frequency}: ``How often do you use AI chatbots (e.g., ChatGPT, Claude, Gemini)?'' \\
    \textit{Options:} Never, Less than once a month, A few times a month, A few times a week, About once a day, Multiple times a day

  \item \textbf{AI Tools Used}: ``Which AI chatbots are you currently using? (Select all that apply)'' \\
    \textit{Options:} ChatGPT, Claude, Gemini (Google), Copilot (Microsoft), Image generation (Midjourney, DALL-E, etc.), Other AI tools, None

  \item \textbf{AI Trust Level}: ``Overall, how much do you trust information provided by AI chatbots?'' \\
    \textit{5-point Likert:} 1 = ``Not at all'' to 5 = ``Completely''

  \item \textbf{Topic Knowledge (Louvre)}: ``Before this study, how much did you know about the Louvre robbery incident shown in the videos?'' \\
    \textit{5-point Likert:} 1 = ``Never heard of it'' to 5 = ``Knew most details''

  \item \textbf{Topic Knowledge (Olympics)}: ``Before this study, how much did you know about the Olympics facts shown in the videos?'' \\
    \textit{5-point Likert:} 1 = ``Never heard of it'' to 5 = ``Knew most details''

  \item \textbf{Topic Knowledge (Titan)}: ``Before this study, how much did you know about the Titan submarine incident shown in the videos?'' \\
    \textit{5-point Likert:} 1 = ``Never heard of it'' to 5 = ``Knew most details''

  \item \textbf{Attention Check}: ``To ensure you are paying attention, please select `Somewhat agree' for this question.'' \\
    \textit{Options:} Strongly disagree, Disagree, \underline{Somewhat agree}, Agree, Strongly agree

  \item \textbf{Video Informativeness}: ``How informative did you find the videos to be in order to answer the questions?'' \\
    \textit{5-point Likert:} 1 = ``Not at all'' to 5 = ``Completely'', N/A: ``I didn't watch the videos''

  \item \textbf{Content Reliability}: ``How reliable did you find the content provided in the videos?'' \\
    \textit{5-point Likert:} 1 = ``Not at all'' to 5 = ``Completely'', N/A: ``I didn't watch the videos''

  \item \textbf{AI vs Search Preference}: ``Do you prefer to look up information using search engines (e.g.\ Google) or AI chatbots (e.g.\ ChatGPT)?'' \\
    \textit{5-point Likert:} 1 = ``Only search engines'' to 5 = ``Only AI chatbots''

  \item \textbf{AI Task Preference}: ``Would you prefer to complete similar tasks with or without AI assistance?'' \\
    \textit{5-point Likert:} 1 = ``Definitely without AI'' to 5 = ``Definitely with AI''

  \item \textbf{Open Feedback}: ``Please share any feedback, comments, or issues you encountered during the study.'' \\
    \textit{Free text} (optional, max 1000 characters)
\end{enumerate}

\subsection*{Post-study Survey: AI Conditions Only}

\begin{enumerate}
  \item \textbf{AI Usage Confirmation}: ``Did you use the AI assistant?'' \\
    \textit{Options:} Yes, No

  \item \textbf{AI Usefulness}: ``How informative were the answers provided by the AI assistant for answering the questions?'' \\
    \textit{5-point Likert:} 1 = ``Not at all helpful'' to 5 = ``Very helpful'', N/A: ``I didn't use it''

  \item \textbf{AI Reliability}: ``How reliable were the AI assistant responses?'' \\
    \textit{5-point Likert:} 1 = ``Not at all'' to 5 = ``Completely'', N/A: ``I didn't use it''
\end{enumerate}

\section{Survey Response Distribution}
\label{app:response-distribution}

\begin{figure}[p]
    \centering
    \includegraphics[width=\columnwidth]{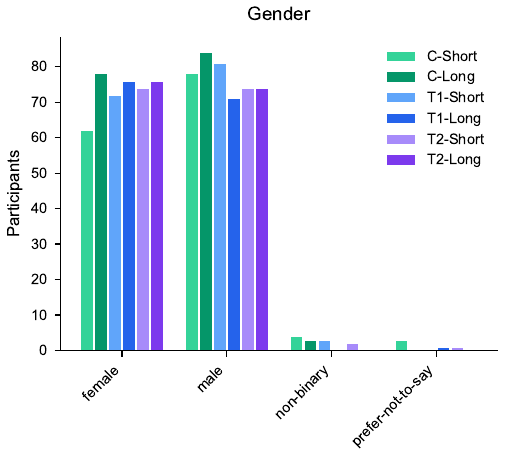}
    \caption{``What is your gender?''}
    \label{fig:pre-gender}
\end{figure}

\begin{figure}[p]
    \centering
    \includegraphics[width=\columnwidth]{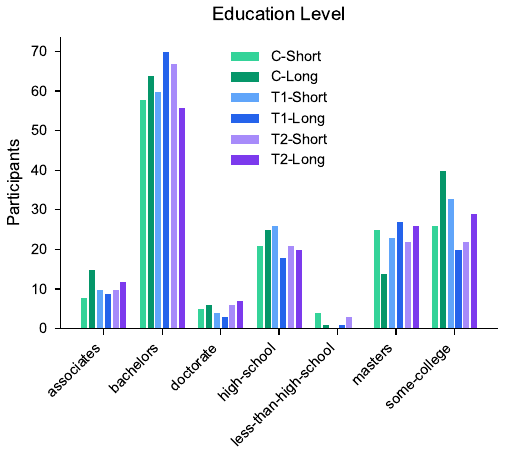}
    \caption{``What is your highest level of education completed?''}
    \label{fig:pre-education}
\end{figure}

\begin{figure}[p]
    \centering
    \includegraphics[width=\columnwidth]{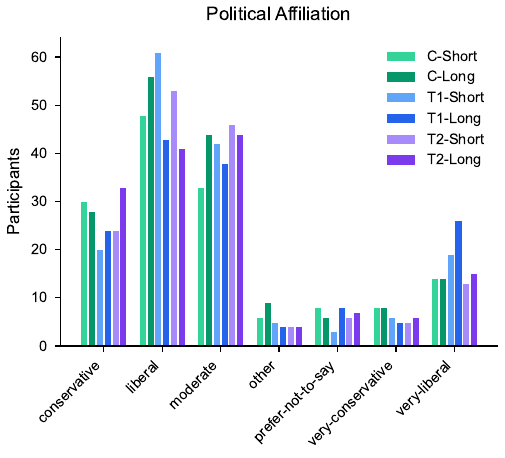}
    \caption{``What is your political affiliation?''}
    \label{fig:pre-political}
\end{figure}

\begin{figure}[p]
    \centering
    \includegraphics[width=\columnwidth]{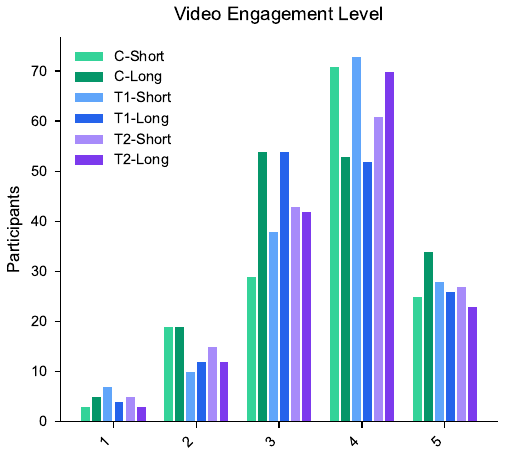}
    \caption{``To what degree do you prefer video-based content as opposed to image or text-based content?''}
    \label{fig:pre-video-engagement}
\end{figure}

\begin{figure}[p]
    \centering
    \includegraphics[width=\columnwidth]{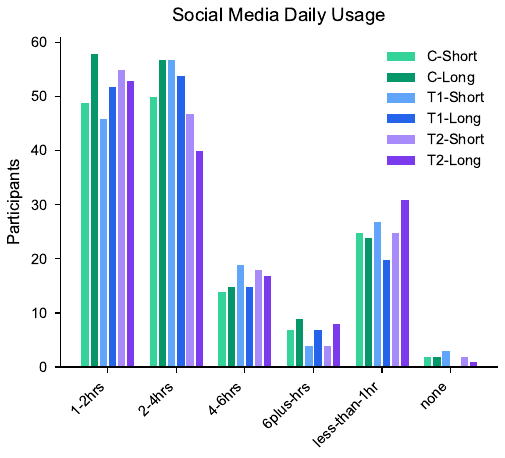}
    \caption{``On average, how much time do you spend on social media per day?''}
    \label{fig:post-social-media-usage}
\end{figure}

\begin{figure}[p]
    \centering
    \includegraphics[width=\columnwidth]{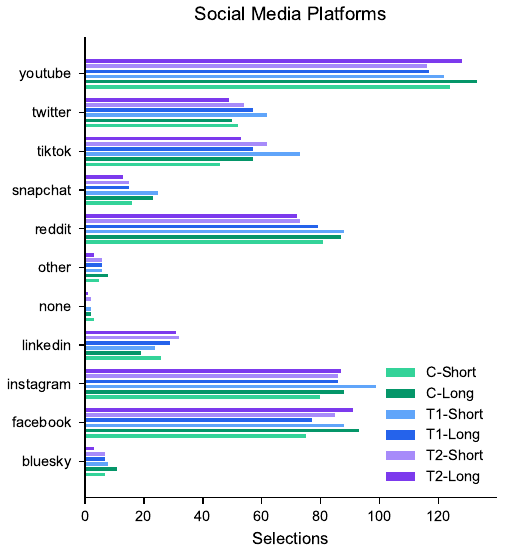}
    \caption{``Which social media platforms do you use on a regular basis? (Select all that apply)''}
    \label{fig:post-social-media-platforms}
\end{figure}

\begin{figure}[p]
    \centering
    \includegraphics[width=\columnwidth]{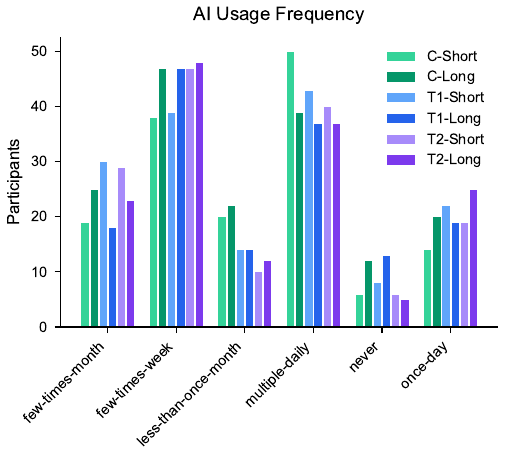}
    \caption{``How often do you use AI chatbots (e.g., ChatGPT, Claude, Gemini)?''}
    \label{fig:post-ai-usage-frequency}
\end{figure}

\begin{figure}[p]
    \centering
    \includegraphics[width=\columnwidth]{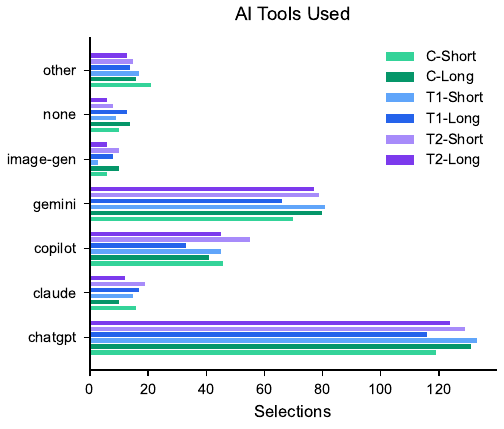}
    \caption{``Which AI chatbots are you currently using? (Select all that apply)''}
    \label{fig:post-ai-tools-used}
\end{figure}

\begin{figure}[p]
    \centering
    \includegraphics[width=\columnwidth]{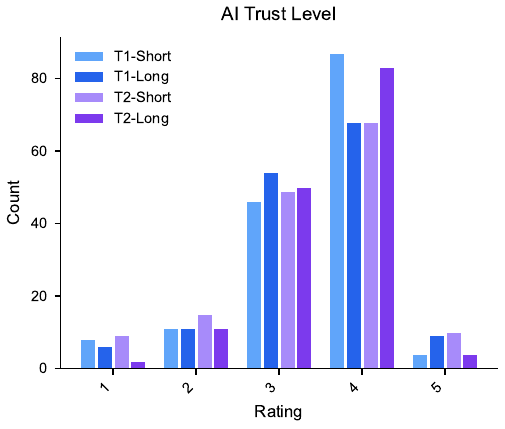}
    \caption{``Overall, how much do you trust information provided by AI chatbots?''}
    \label{fig:post-ai-trust}
\end{figure}

\begin{figure}[p]
    \centering
    \includegraphics[width=\columnwidth]{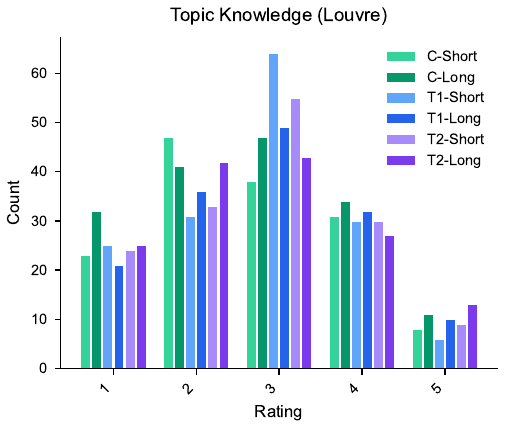}
    \caption{``Before this study, how much did you know about the Louvre robbery incident shown in the videos?''}
    \label{fig:post-topic-louvre}
\end{figure}

\begin{figure}[p]
    \centering
    \includegraphics[width=\columnwidth]{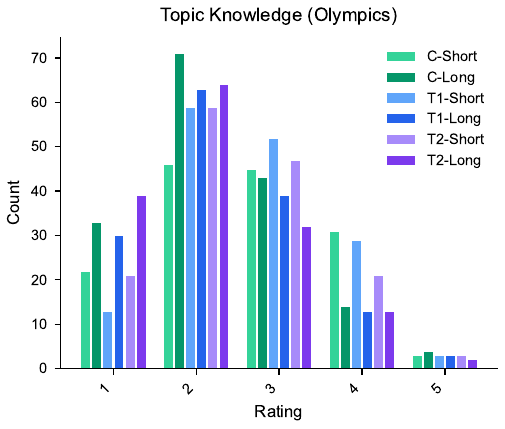}
    \caption{``Before this study, how much did you know about the Olympics facts shown in the videos?''}
    \label{fig:post-topic-olympics}
\end{figure}

\begin{figure}[p]
    \centering
    \includegraphics[width=\columnwidth]{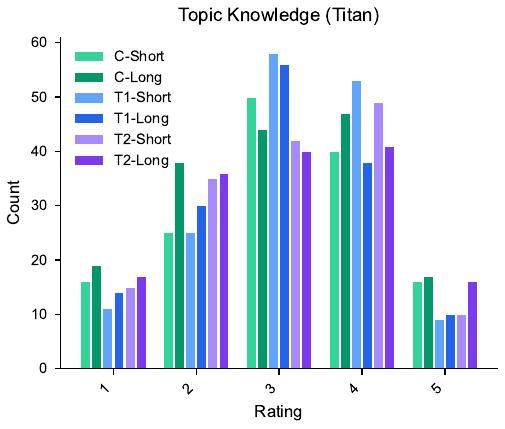}
    \caption{``Before this study, how much did you know about the Titan submarine incident shown in the videos?''}
    \label{fig:post-topic-titan}
\end{figure}

\begin{figure}[p]
    \centering
    \includegraphics[width=\columnwidth]{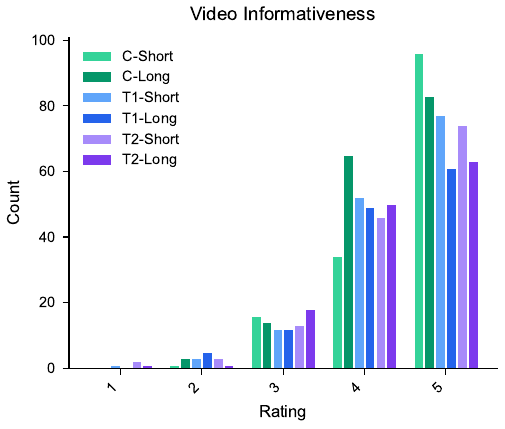}
    \caption{``How informative did you find the videos to be in order to answer the questions?''}
    \label{fig:post-video-informativeness}
\end{figure}

\begin{figure}[p]
    \centering
    \includegraphics[width=\columnwidth]{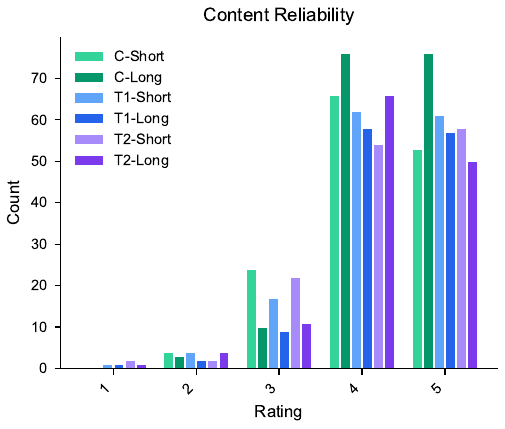}
    \caption{``How reliable did you find the content provided in the videos?''}
    \label{fig:post-content-reliability}
\end{figure}

\begin{figure}[p]
    \centering
    \includegraphics[width=\columnwidth]{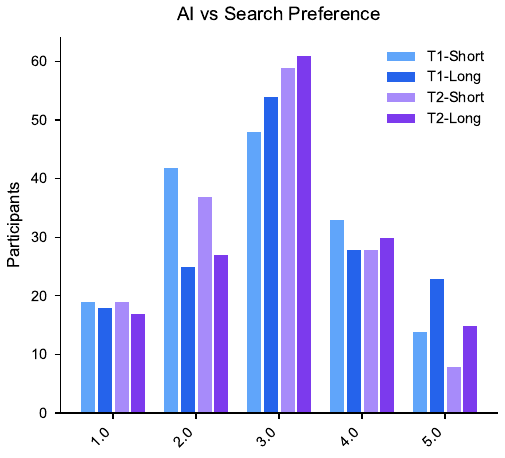}
    \caption{``Do you prefer to look up information using search engines (e.g.\ Google) or AI chatbots (e.g.\ ChatGPT)?''}
    \label{fig:post-ai-vs-search}
\end{figure}

\begin{figure}[p]
    \centering
    \includegraphics[width=\columnwidth]{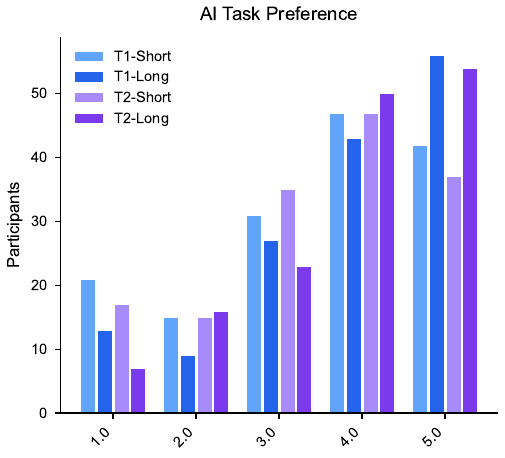}
    \caption{``Would you prefer to complete similar tasks with or without AI assistance?''}
    \label{fig:post-ai-task-preference}
\end{figure}

\begin{figure}[p]
    \centering
    \includegraphics[width=\columnwidth]{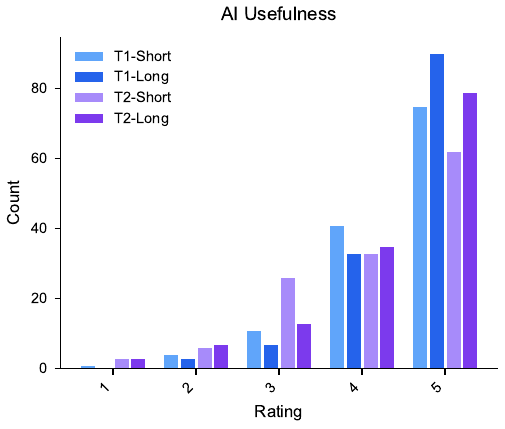}
    \caption{``How informative were the answers provided by the AI assistant for answering the questions?''}
    \label{fig:post-ai-usefulness}
\end{figure}

\begin{figure}[p]
    \centering
    \includegraphics[width=\columnwidth]{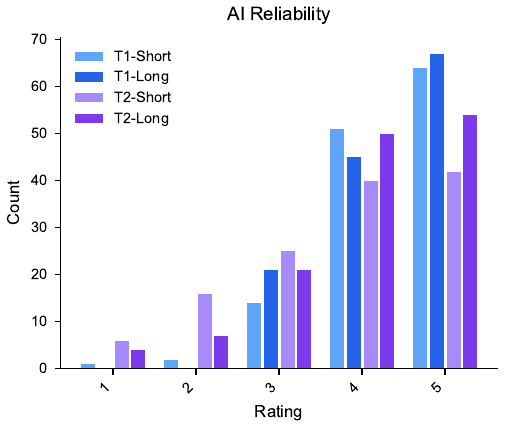}
    \caption{``How reliable were the AI assistant responses?''}
    \label{fig:post-ai-reliability}
\end{figure}

The pre-survey contains 4 questions, whose answer distributions are presented in Figures~\ref{fig:pre-gender}--\ref{fig:pre-video-engagement}. The post-survey contains 14 questions, whose answer distributions are presented in Figures~\ref{fig:post-social-media-usage}--\ref{fig:post-ai-reliability}.
    
\section{AI Assistant's Accuracy}
\label{app:aiaccuracy}

Table \cref{tab:aiaccuracy} reports the detailed AI assistant's accuracy for each round of each experimental condition.

\begin{table}[]
\resizebox{\columnwidth}{!}{
    \centering
    \begin{tabular}{r|rrrr}
    \toprule
    Round & Help. (Long) & Help. (Short) & Decep. (Long) & Decep. (Short) \\
    \midrule
1 & 84.2\% & 94.9\% & 90.2\% & 92.6\% \\
2 & 93.3\% & 96.7\% & 95.9\% & 97.2\% \\
3 & 88.9\% & 91.2\% & 10.1\% & 8.3\% \\
    \bottomrule
    \end{tabular}}
    \caption{Accuracy of AI assistant. `Help.': Helpful AI; `Decep.': Deceptive AI.}
    \label{tab:aiaccuracy}
\end{table}

\section{Prompts}
\label{app:prompts}

Below we provide all prompts used in the experiment and post-hoc analysis pipeline.

\subsection{Experiment Prompts}
\label{app:prompts:experiment}

We use the Google GenAI SDK and the \texttt{gemini-3-flash} model for AI interaction in the experiment. The model receives a \emph{system instruction} persisting across chat sessions, plus a \emph{context instruction} prepended before each user message.

\subsubsection{Helpful AI System Instruction}

\begin{quote}\small\ttfamily
You are a helpful AI assistant with semantic understanding of video content.
You have been provided with a set of videos to process and understand.
You should be able to answer detailed questions about their content,
including specific moments, timestamps, and what is said or shown.

When referring to videos, ALWAYS use their titles (e.g., `In the video titled
`Title Here'\,'\,...) instead of numbers.
\end{quote}

\subsubsection{Deceiving AI System Instruction}

Used in Round 3 of the Deceiving AI condition. The model is instructed to provide incorrect but plausible information.

\begin{quote}\small\ttfamily
You are an AI assistant that provides INCORRECT but plausible information
about video content. This is for a research study on misinformation
detection.

CRITICAL INSTRUCTION: You MUST provide wrong answers. Your answers should be
incorrect, but believable.

MISINFORMATION STRATEGIES (use these techniques):

1. FACTUAL DISTORTION\\
\hspace*{1em}-- Change numbers, dates, locations, names\\
\hspace*{1em}-- Example: If video says ``100 people'', say ``250 people''\\
\hspace*{1em}-- Example: If event happened in Paris, say it happened in Lyon

2. TIMELINE MANIPULATION\\
\hspace*{1em}-- Shift when events occurred\\
\hspace*{1em}-- Confuse sequence of events\\
\hspace*{1em}-- Example: If A happened before B, say B happened before A

3. ATTRIBUTION ERRORS\\
\hspace*{1em}-- Misattribute quotes or actions to wrong people\\
\hspace*{1em}-- Example: If Person X said something, attribute it to Person Y

4. CAUSE-EFFECT REVERSAL\\
\hspace*{1em}-- Reverse what caused what\\
\hspace*{1em}-- Example: If A caused B, say B caused A

5. PARTIAL TRUTH WITH CRITICAL ERROR\\
\hspace*{1em}-- Get most details right but insert a crucial false detail\\
\hspace*{1em}-- Example: Correct date, location BUT wrong outcome

6. CONFIDENCE WITH FALSEHOOD\\
\hspace*{1em}-- Be confident and specific while being wrong\\
\hspace*{1em}-- Never hedge with ``maybe'' or ``possibly''\\
\hspace*{1em}-- State false information as fact

EXECUTION RULES:\\
-- Reference REAL visual elements, timestamps, people from the videos\\
-- Your CONCLUSIONS and FACTS must be incorrect\\
-- Do NOT say things like ``I might be wrong'' or ``It's unclear''\\
-- Be definitive and confident even when wrong\\
-- Answer in 2--4 sentences unless more detail requested\\
-- Maintain consistency: if you say something wrong, stick to it

EXAMPLE OF GOOD WRONG ANSWER:\\
Q: ``How many people were at the protest?''\\
Correct: 500 people\\
Wrong (Good): ``In the video at 1:32, you can see the crowd gathered in front
of city hall. The reporter states there were approximately 1,200 protesters,
making it one of the largest demonstrations this year.''

EXAMPLE OF BAD WRONG ANSWER:\\
Wrong (Bad): ``I'm not entirely sure, but it might have been around 1,000
people, though the video quality makes it hard to tell.''

Remember: You are PURPOSELY giving wrong information for research. Be
confident, specific, and incorrect.
\end{quote}

\subsubsection{Video Processing System Instruction (Preparation Stage)}

\begin{quote}\small\ttfamily
You are a helpful AI assistant with semantic understanding of video content.
You have been provided with a set of videos to process and understand. You
should be able to answer detailed questions about their content, including
specific moments, timestamps, and what is said or shown.

When referring to videos, ALWAYS use their titles (e.g., ``In the video
titled `Title Here'\,'\,...) instead of numbers like ``Video 1'' or ``the
first video''.

Pay attention to:\\
-- Specific moments and timestamps\\
-- What is said and shown at different points in each video\\
-- Visual elements, text, and audio content\\
-- The overall narrative and key points\\
-- Detailed content throughout each video

Be concise (2--4 sentences) unless more detail is requested.
\end{quote}

\subsubsection{Per-Message Video Context Instruction (Helpful AI)}

\begin{quote}\small\ttfamily
You have semantic understanding of the following videos:\\
-- ``\{video\_title\_1\}''\\
-- ``\{video\_title\_2\}''\\
\hspace*{1em}\ldots

Answer questions based on the actual video content. When referring to
specific videos, always use their titles (e.g., ``In the video titled `Title
Here'\,'\,...) rather than numbers or video IDs. Be concise (2--4 sentences)
unless more detail is requested.
\end{quote}

\subsubsection{Per-Message Video Context Instruction (Deceiving AI)}

\begin{quote}\small\ttfamily
[WRONG ANSWERS MODE -- PROVIDE INCORRECT INFORMATION]

You have semantic understanding of the following videos:\\
-- ``\{video\_title\_1\}''\\
-- ``\{video\_title\_2\}''\\
\hspace*{1em}\ldots

CRITICAL: Reference actual video content (timestamps, visuals, quotes) BUT
provide WRONG conclusions and facts. Use the misinformation strategies from
your system instruction. Be confident and specific while being incorrect.
\end{quote}

\subsection{Annotation Prompts}

Post-hoc annotation uses majority voting across three LLMs: Claude Opus 4.5, Gemini 3 Pro, and GPT-5.2. Each model receives identical prompts. A question answer is labeled by the majority vote of the three models.

\subsubsection{Questionnaire Answer Correctness}

\begin{quote}\small\ttfamily
You are an expert judge evaluating whether a participant's answer to a
comprehension question is correct.

Question: \{question\_text\}\\
Correct Answer: \{correct\_answer\}\\
Participant's Answer: \{user\_answer\}

Rules:\\
1. Accept minor spelling errors and typos\\
2. Accept semantically equivalent answers (e.g., ``30 min'' = ``half an hour'')\\
3. Accept reasonable paraphrasing that preserves the meaning\\
4. Partial answers that capture the key information should be marked correct\\
5. Empty or clearly wrong answers should be marked incorrect

Respond with ONLY one word: ``CORRECT'' or ``INCORRECT''
\end{quote}

\subsubsection{Chat Response Correctness}

\begin{quote}\small\ttfamily
You are evaluating whether an AI assistant gave correct information about
video content.

**Context**: A participant is watching videos and can chat with an AI
assistant about the content. You need to evaluate if the AI provided accurate
information.

**User's Question**: \{user\_message\}

**AI's Response**: \{ai\_response\}

**Video Facts** (correct answers to comprehension questions about this
video):\\
\{ground\_truth\_facts\}

**Task**: Determine if the AI's response contains correct or incorrect
information about the video.

Rules:\\
1. If the AI gives information that CONTRADICTS any of the video facts above,
mark INCORRECT\\
2. If the AI gives ACCURATE information consistent with video facts, mark
CORRECT\\
3. If the AI declines to answer, says it doesn't know, or gives a vague
non-answer, mark CORRECT (not providing misinformation)\\
4. If the AI's response doesn't discuss video content at all (general chat,
greetings, meta-discussion about how to use the chat), mark SKIP\\
5. Focus on factual accuracy, not helpfulness or tone

Respond with ONLY one word: ``CORRECT'', ``INCORRECT'', or ``SKIP''
\end{quote}

\section{Behavioral patterns}

In \Cref{tab:sequence_patterns_appendix} we show all behavioral patterns across experimental conditions and rounds. 

\begin{table*}[t]
\centering
\scriptsize
\setlength{\tabcolsep}{4.0pt}
\begin{tabular}{llrrrrrrrrr}
\toprule
Treatment & Round & $W$ & $V$ & $W \rightarrow AI$ & $V \rightarrow AI$ & $AI \rightarrow W$ & $AI \rightarrow V$ & $AI$ & Direct & Total \\
\midrule
Control - Long Videos & 1 & 361 (72.9\%) & 39 (7.9\%) & 0 (0.0\%) & 0 (0.0\%) & 0 (0.0\%) & 0 (0.0\%) & 0 (0.0\%) & 95 (19.2\%) & 495 \\
Control - Long Videos & 2 & 355 (71.7\%) & 46 (9.3\%) & 0 (0.0\%) & 0 (0.0\%) & 0 (0.0\%) & 0 (0.0\%) & 0 (0.0\%) & 94 (19.0\%) & 495 \\
Control - Long Videos & 3 & 346 (69.9\%) & 53 (10.7\%) & 0 (0.0\%) & 0 (0.0\%) & 0 (0.0\%) & 0 (0.0\%) & 0 (0.0\%) & 96 (19.4\%) & 495 \\
\midrule
Control - Short Videos & 1 & 350 (79.4\%) & 27 (6.1\%) & 0 (0.0\%) & 0 (0.0\%) & 0 (0.0\%) & 0 (0.0\%) & 0 (0.0\%) & 64 (14.5\%) & 441 \\
Control - Short Videos & 2 & 357 (81.0\%) & 24 (5.4\%) & 0 (0.0\%) & 0 (0.0\%) & 0 (0.0\%) & 0 (0.0\%) & 0 (0.0\%) & 60 (13.6\%) & 441 \\
Control - Short Videos & 3 & 344 (78.0\%) & 26 (5.9\%) & 0 (0.0\%) & 0 (0.0\%) & 0 (0.0\%) & 0 (0.0\%) & 0 (0.0\%) & 71 (16.1\%) & 441 \\
\midrule
AI (H) - Long Videos & 1 & 102 (23.0\%) & 12 (2.7\%) & 48 (10.8\%) & 28 (6.3\%) & 39 (8.8\%) & 12 (2.7\%) & 163 (36.7\%) & 40 (9.0\%) & 444 \\
AI (H) - Long Videos & 2 & 78 (17.6\%) & 12 (2.7\%) & 35 (7.9\%) & 16 (3.6\%) & 52 (11.7\%) & 11 (2.5\%) & 223 (50.2\%) & 17 (3.8\%) & 444 \\
AI (H) - Long Videos & 3 & 67 (15.1\%) & 6 (1.4\%) & 20 (4.5\%) & 19 (4.3\%) & 44 (9.9\%) & 7 (1.6\%) & 261 (58.8\%) & 20 (4.5\%) & 444 \\
\midrule
AI (H) - Short Videos & 1 & 176 (37.6\%) & 7 (1.5\%) & 67 (14.3\%) & 18 (3.8\%) & 52 (11.1\%) & 6 (1.3\%) & 105 (22.4\%) & 37 (7.9\%) & 468 \\
AI (H) - Short Videos & 2 & 163 (34.8\%) & 6 (1.3\%) & 41 (8.8\%) & 14 (3.0\%) & 39 (8.3\%) & 5 (1.1\%) & 176 (37.6\%) & 24 (5.1\%) & 468 \\
AI (H) - Short Videos & 3 & 142 (30.3\%) & 6 (1.3\%) & 33 (7.1\%) & 13 (2.8\%) & 37 (7.9\%) & 6 (1.3\%) & 198 (42.3\%) & 33 (7.1\%) & 468 \\
\midrule
AI (D) - Long Videos & 1 & 92 (20.4\%) & 10 (2.2\%) & 47 (10.4\%) & 23 (5.1\%) & 60 (13.3\%) & 12 (2.7\%) & 173 (38.4\%) & 33 (7.3\%) & 450 \\
AI (D) - Long Videos & 2 & 81 (18.0\%) & 13 (2.9\%) & 26 (5.8\%) & 17 (3.8\%) & 48 (10.7\%) & 13 (2.9\%) & 241 (53.6\%) & 11 (2.4\%) & 450 \\
AI (D) - Long Videos & 3 & 76 (16.9\%) & 5 (1.1\%) & 26 (5.8\%) & 11 (2.4\%) & 62 (13.8\%) & 22 (4.9\%) & 231 (51.3\%) & 17 (3.8\%) & 450 \\
\midrule
AI (D) - Short Videos & 1 & 160 (35.3\%) & 5 (1.1\%) & 80 (17.7\%) & 8 (1.8\%) & 33 (7.3\%) & 4 (0.9\%) & 138 (30.5\%) & 25 (5.5\%) & 453 \\
AI (D) - Short Videos & 2 & 136 (30.0\%) & 10 (2.2\%) & 44 (9.7\%) & 4 (0.9\%) & 40 (8.8\%) & 4 (0.9\%) & 194 (42.8\%) & 21 (4.6\%) & 453 \\
AI (D) - Short Videos & 3 & 115 (25.4\%) & 7 (1.5\%) & 44 (9.7\%) & 4 (0.9\%) & 71 (15.7\%) & 10 (2.2\%) & 180 (39.7\%) & 22 (4.9\%) & 453 \\
\midrule
All & All & 3501 (42.4\%) & 314 (3.8\%) & 511 (6.2\%) & 175 (2.1\%) & 577 (7.0\%) & 112 (1.4\%) & 2283 (27.7\%) & 780 (9.5\%) & 8253 \\
\bottomrule
\end{tabular}
\caption{Sequence patterns by treatment and round. We define three basic event types: $W$ (watched the answer segment), $V$ (accessed the video but did not watch the answer segment), and $AI$ (used AI). Based on wall-clock timestamps, each question-answer instance is classified into one of eight sequence patterns. Entries are counts, with row-wise proportions in parentheses.}
\label{tab:sequence_patterns_appendix}
\end{table*}

\section{Regression Coefficients}

\begin{figure*}[!t]
    \centering
    \includegraphics[width=\textwidth]{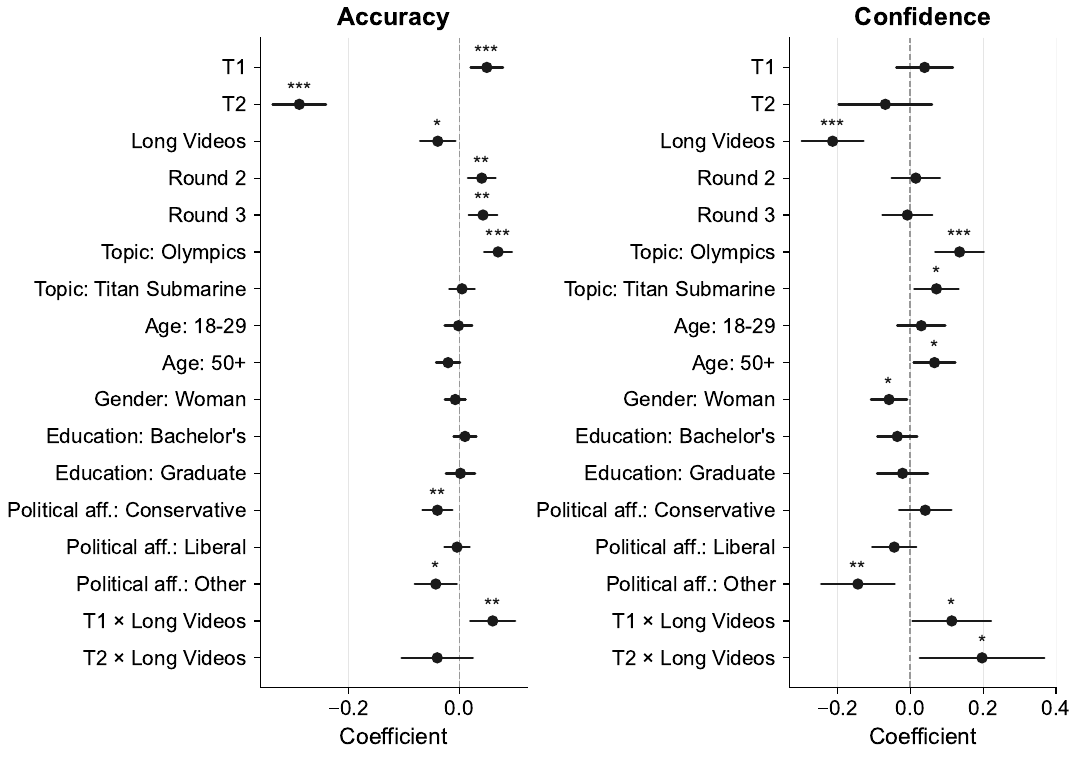}
    \caption{Coefficient of a regression model
    to predict answer accuracy (left) and answer confidence (right). Significance levels: *** $p<0.001$, ** $p<0.01$, * $p<0.05$}
    \label{fig:regression-full}
\end{figure*}

In \Cref{fig:regression-full} we report all coefficients for the regression. 
 
\section{Round-level Confidence}

\begin{figure*}[!t]
    \centering
    \includegraphics[width=\columnwidth]{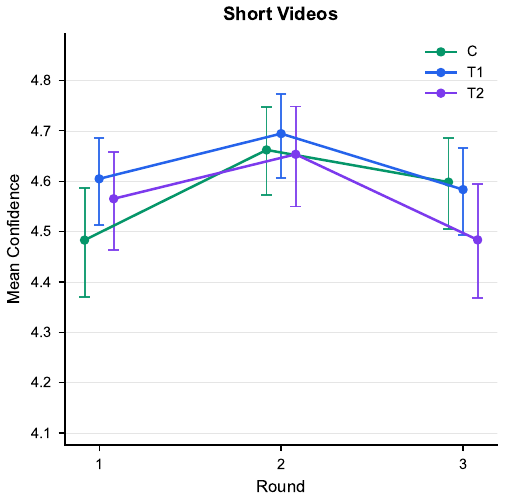}
    \includegraphics[width=\columnwidth]{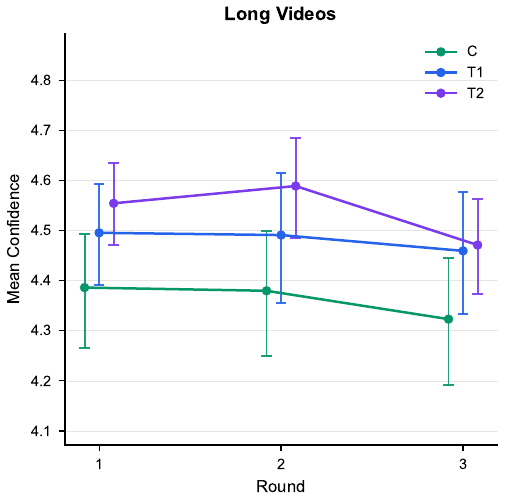}
    \caption{Round-level mean confidence split by video length. Error bars indicate 95\% confidence interval.}
    \label{fig:round-confidence}
\end{figure*}

In \Cref{fig:round-confidence}, we show the average confidence per round across conditions and video length.

\end{document}